\documentclass[12pt]{article}
\usepackage{graphicx}
\DeclareGraphicsExtensions{.pdf,.jpg,.jpeg}
\newcommand\pubdate{\today}

\def\napoli{University of Oxford, \\ Denys Wilkinson Building, Keble Road, Oxford, OX1 3RH,  UK}

\def\Title#1{\begin{center} {\Large #1 } \end{center}}
\def\Author#1{\begin{center}{ \sc #1} \end{center}}
\def\Address#1{\begin{center}{ \it #1} \end{center}}

\newcommand\pubblock{\rightline{\begin{tabular}{l} 
\\
         \pubdate  \end{tabular}}}
\newenvironment{Abstract}{\begin{quotation}  }{\end{quotation}}
\newenvironment{Presented}{\begin{quotation} \begin{center} 
             PRESENTED AT\end{center}\bigskip 
      \begin{center}\begin{large}}{\end{large}\end{center} \end{quotation}}
\def\Acknowledgements{\bigskip  \bigskip \begin{center} \begin{large}
             \bf ACKNOWLEDGEMENTS \end{large}\end{center}}

\def\beq{\begin{equation}}
\def\eeq#1{\label{#1}\end{equation}}
\def\eeqn{\end{equation}}

\def\beqa{\begin{eqnarray}}
\def\eeqa#1{\label{#1}\end{eqnarray}}
\def\eeqan{\end{eqnarray}}

\let\bar=\overbar

\def\Dslash{\not{\hbox{\kern-4pt $D$}}}
\def\dslash{\not{\hbox{\kern-2pt $\del$}}}

\def\msb{{\bar{\ssstyle M \kern -1pt S}}}

\begin{document}
\begin{titlepage}
\pubblock

\vfill
\Title{Dawn or dusk?  Flavour physics in the hadron collider era}
\vfill
\Author{ Guy Wilkinson}
\Address{\napoli}
\vfill
\begin{Abstract}
A review is made of the status of, and prospects for,  flavour physics studies at hadron colliders 
in the `post $e^+ e^-$ era'.  It is argued that exciting times lie ahead.
\end{Abstract}
\vfill
\begin{Presented}
 CKM 2010, the 6th International Workshop on the CKM \\ \vspace*{0.1cm} Unitarity Triangle, University of Warwick, UK, 6-10/9/2010
\end{Presented}
\vfill
\end{titlepage}
\def\thefootnote{\fnsymbol{footnote}}
\setcounter{footnote}{0}

\section{Onset of the Dark Ages}

In the heavy-flavour community at present, it is not uncommon to detect sentiments similar to those
expressed by Petrarch in the early fourteenth century:

\vspace*{0.2cm}

{\it ``My fate is to live among varied and confusing storms.  But for you perhaps, if as I hope and wish you
will live long after me, there will follow a better age.  This sleep of forgetfulness will not last for ever. When
the darkness has been dispersed, our descendants can come again in the former pure radiance.''}

\vspace*{0.2cm}

Here the writer bemoans the end of the classical age and his despair at living in a time of ignorance 
and low civilisation -- the so-called Dark Ages.    Many attendees at this workshop will know how he felt,  for it
is the first CKM meeting of the post-$B$ factory era, with neither BABAR or Belle any longer taking data.
In the Dark Ages, the pursuit of new knowledge ceased.  Instead lone scholars in remote monasteries
worked tirelessly on recording all that was known, so that existing learning would not be lost to future
generations.  So it is now, where efforts are underway to produce documents  such as the `$B$-factory Legacy 
Book'~\cite{BFLB}.   Of course, after the Dark Ages came the Renaissance, and so physicists are comforted
with the knowledge that the resumption of the classical ways (`former pure radiance') will arrive in
due course -- and indeed the $e^+e^-$ programme will be reborn with the Super-$B$ and Belle II projects.

A further unpleasant feature of the Dark Ages was the threat from barbarian hordes: brutish,
uncivilised and terrible new forces which roamed unchecked.  It is maybe unfair to accuse members
of the flavour-community of regarding the LHC in a such terms, but the `brute force' approach 
of direct observation is certainly different to that which was employed at the $B$-factories.
Of course, with LHCb, the new machine has a dedicated flavour-physics experiment; furthermore, ATLAS and CMS
themselves have goals in the flavour sector. Nevertheless, doubt is sometimes
expressed that the harsh  hadronic environment will allow for measurements to be performed that
are of the same quality and interest of those that are possible at $e^+e^-$ machines.

An alternative view, notwithstanding the remarkable achievements of BABAR and Belle, is that flavour physics
is now embarked on a new golden age.  It should be recalled that historically hadron colliders have made 
important contributions to this discipline, for example with the observation of $B$-mixing~\cite{UA1MIX}.
The dawn of the present era of enlightenment can be dated to the Tevatron's observation of
$B^0_s$ oscillations in 2006~\cite{CDFMIX}, and since then CDF and D0  have produced many flavour
results of outstanding interest, with a very large amount of data still to be analysed.  The successful 
start-up to the LHC gives hope that this programme will continue at CERN. Indeed -- as will
be argued in this report -- it is not unlikely that flavour physics will provide the headline measurements
of the 2010-12  LHC run.

\section{Machine status and near-term prospects}

The ongoing Tevatron `Run II' will continue until the end of 2011.  Extrapolations of the performance achieved
until now suggest that around $12~\,{\rm fb^{-1}}$ per experiment in total could be delivered on this timescale.
Such a sample would constitute two to three times more data than have been used in analyses presented until now,
and would result in a corresponding improvement in sensitivity for many interesting flavour-physics topics (not to mention
the Higgs search, which is clearly beyond the remit of this review).  

At the time of the workshop the LHC had delivered around $3-4~\rm {\rm pb^{-1}}$ of data  per experiment
at $\sqrt{s}=7\,{\rm TeV}$~\footnote{ALICE 
recorded much less data, as it is designed to operate at  a much lower instantaneous luminosity than the other three experiments.}, 
with a start of
fill luminosity of $\sim 10^{31} \, {\rm cm^{-2} s^{-1}}$.  
In the weeks that followed the number of colliding bunches
was gradually increased, with the result that the luminosity was eventually boosted
to a few $10^{32} \, {\rm cm^{-2} s^{-1}}$.   The total integrated 
luminosity delivered in proton collisions was $\sim 40 \, {\rm pb^{-1}}$ per experiment, most of which was accumulated in a 
two week period at the end of the run.  

The special running conditions of the 2010 run compared with those that are foreseen in future
both brought opportunities for the experiments, and presented challenges.  
The enormous change in instantaneous luminosity from the start to the end of the run meant
that the trigger strategies had to evolve continuously, but thresholds could be placed at rather low values 
early on, with a consequent benefit in efficiency.  In terms of emittance and bunch charge, 
values close to design specifications were
quickly reached. This meant that the `pile-up' per bunch crossing soon became significant.
In contrast to the General Purpose Detectprs (GPDs), LHCb is designed to take data in a low pile-up environment,
and run at a luminosity of a few $10^{32} \,{\rm cm^{-2} s^{-1}}$.  This
luminosity was reached during 2010, but with around 2.5 interactions per crossing
in contrast to the LHCb design value of 0.4.  Operating in these conditions placed
a strain both on the trigger, and -- with the larger than foreseen event sizes -- on the offline
computing.
  
In the 2011-12 LHC run the luminosity at the GPD interaction points will be 
increased, while LHCb will continue to take data in the $10^{32} \,{\rm cm^{-2 }s^{-1}}$
regime, but at this fixed luminosity the pile-up rate will diminish, as the number of bunches
in the machine is increased. It is not
possible to make reliable predictions for the amount of data that will be accumulated,
but estimates of $\sim 1-2 \, {\rm fb^{-1}}$ for LHCb, and several times more for ATLAS
and CMS, are not unreasonable.
 
Most of the numerical results available at the workshop were obtained from the $\sim 10 \, {\rm nb^{-1}}$ of integrated
luminosity that had been analysed for the summer conferences, although plots were available for samples 
of $0.1$ -- $1\, {\rm pb^{-1}}$ of data.
Where possible the plots and results included in this write-up have been updated to larger datasets.

\section{Early heavy flavour results from the LHC}

\begin{figure}[htb]
\centering
\includegraphics[height=0.34\textwidth]{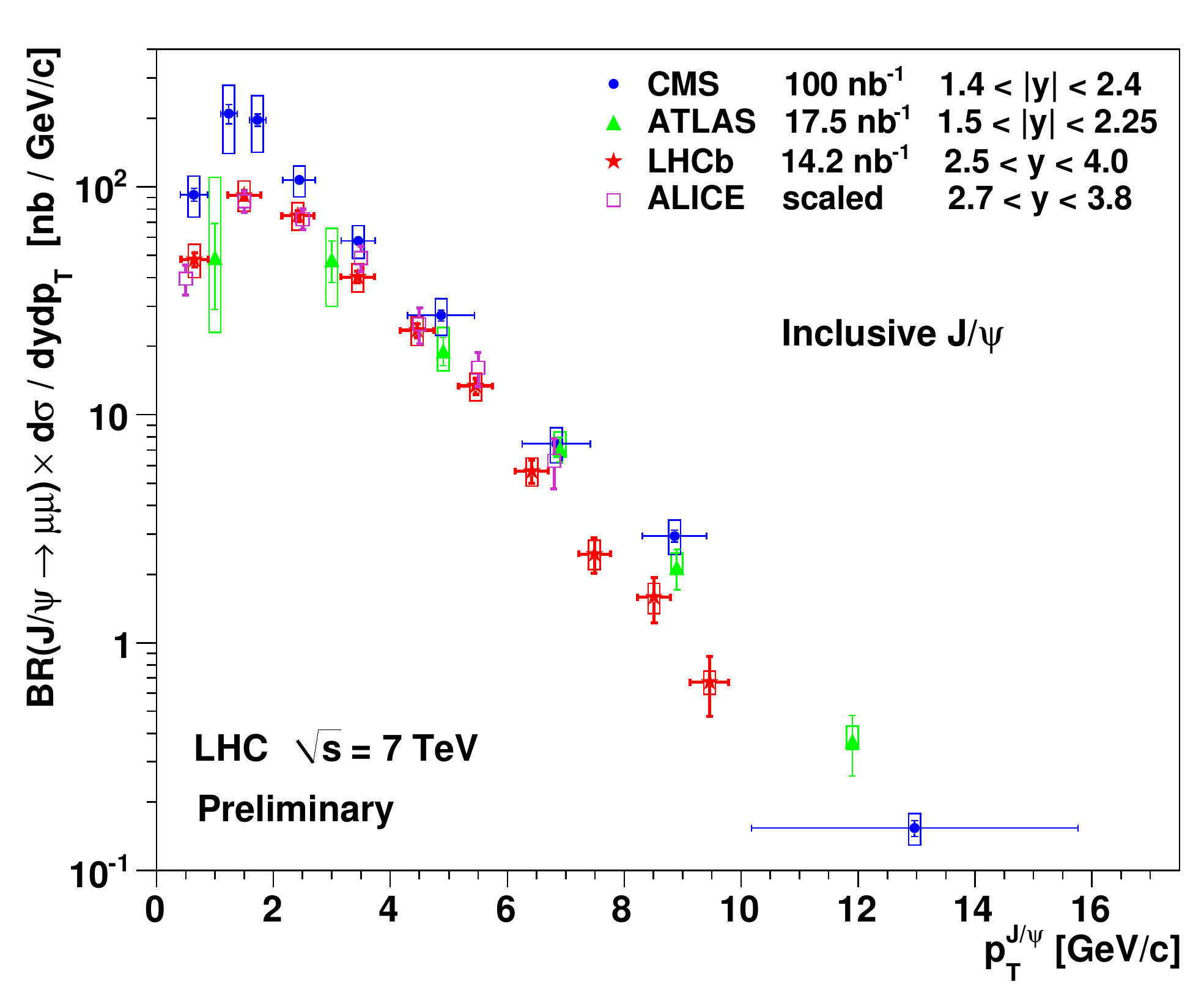} 
\includegraphics[height=0.34\textwidth]{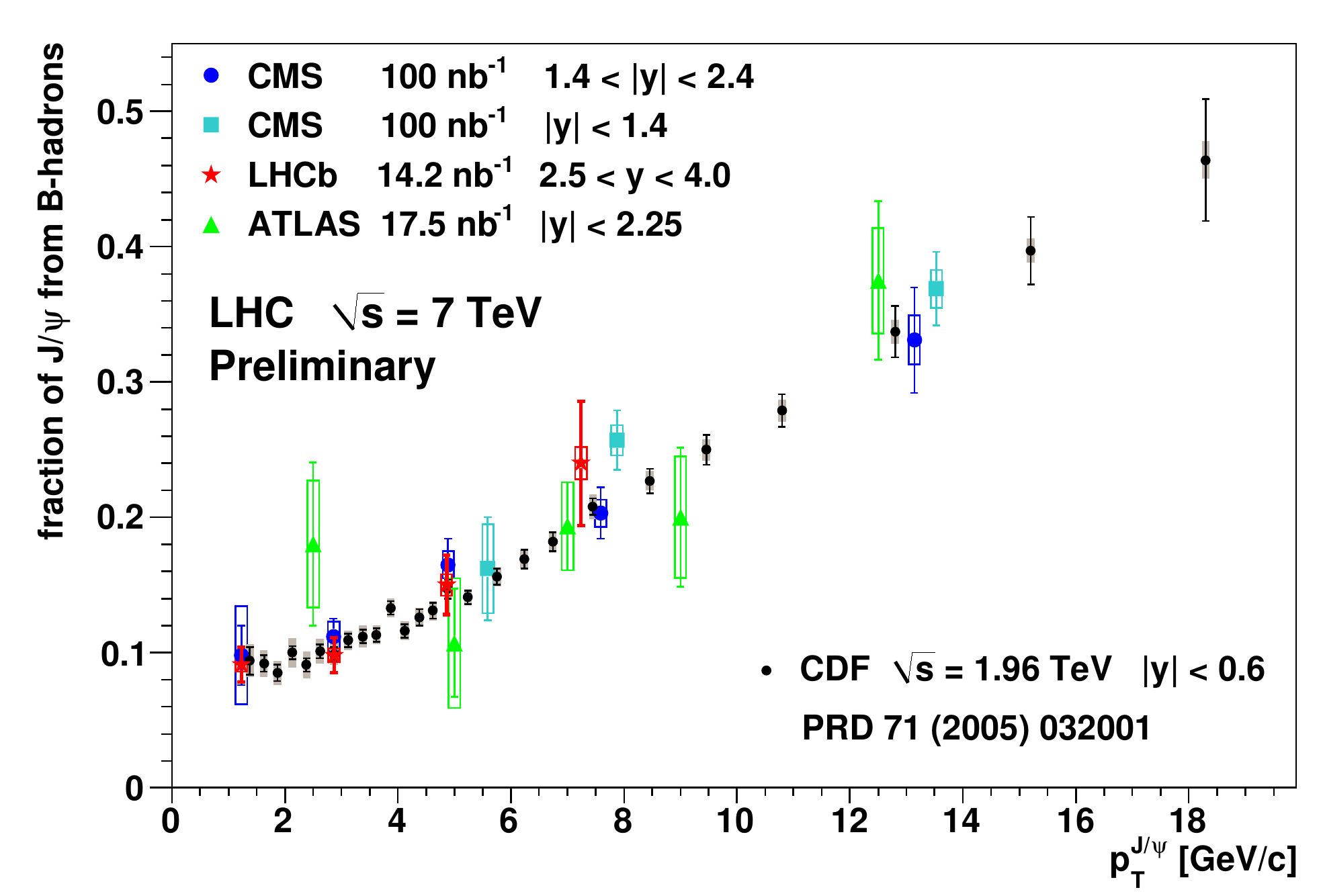}
\caption{
LHC $J/\psi$ results from ICHEP 2010.  
Left: $BR(J/\psi \to \mu\mu) \times d\sigma/ d y d p_T $ vs. $p_T^{J/\psi}$. 
Right: fraction of $J/\psi$ from $B$-hadrons vs. $p_T^{J/\psi}$, also showing the CDF results~\cite{JPSICOMP}.
}
\label{fig:jpsiichep}
\end{figure}

A natural early topic of study with LHC data is quarkonia.  The production mechanism of 
$J/\psi$ mesons and heavier states is not well understood at hadron colliders~\cite{JPSIPUZZLE}.
It is natural therefore to begin to study this topic at the LHC.   In the $c\bar{c}$ system the data from 
the 2010 LHC run will allow for the measurement of inclusive prompt (i.e. not from $B$ decay) $J/\psi$ 
production cross-sections, polarisation studies, and measurements of $\psi(2S)$ and $\chi_c$ production.
Taken together, these results should be able to discriminate between a range of production models.
Already available at the time of the workshop, and in several cases updated since, are measurements
from each collaboration on inclusive production~\cite{CMSJPSI_ICHEP,ATLASJPSI_ICHEP,LHCBJPSI_ICHEP,CMS_JPSI,ALICEJPSI_DEC,LHCBJPSI_DEC}.  
A further attraction of isolating a $J/\psi$ sample
is that decays displaced from the primary interaction vertex are a convenient signature
of $b$-hadron production.  Therefore these data have been used both to measure the inclusive differential $J/\psi$ cross-section,
and to calculate the fraction of $J/\psi$ mesons arising from $b$-hadron decays.  The results presented at the 
ICHEP 2010 conference are plotted  in Fig.~\ref{fig:jpsiichep},
as a function of transverse momentum of the $J/\psi$~\cite{JPSICOMP}.  The shape of the
differential cross-section is very similar for each experiment, despite different acceptances
in rapidity.  The fraction of $J/\psi$ events from $b$-hadron decays agrees well between all LHC experiments,
and the Tevatron.

\begin{figure}[htb]
\centering
\includegraphics[height=0.38\textwidth]{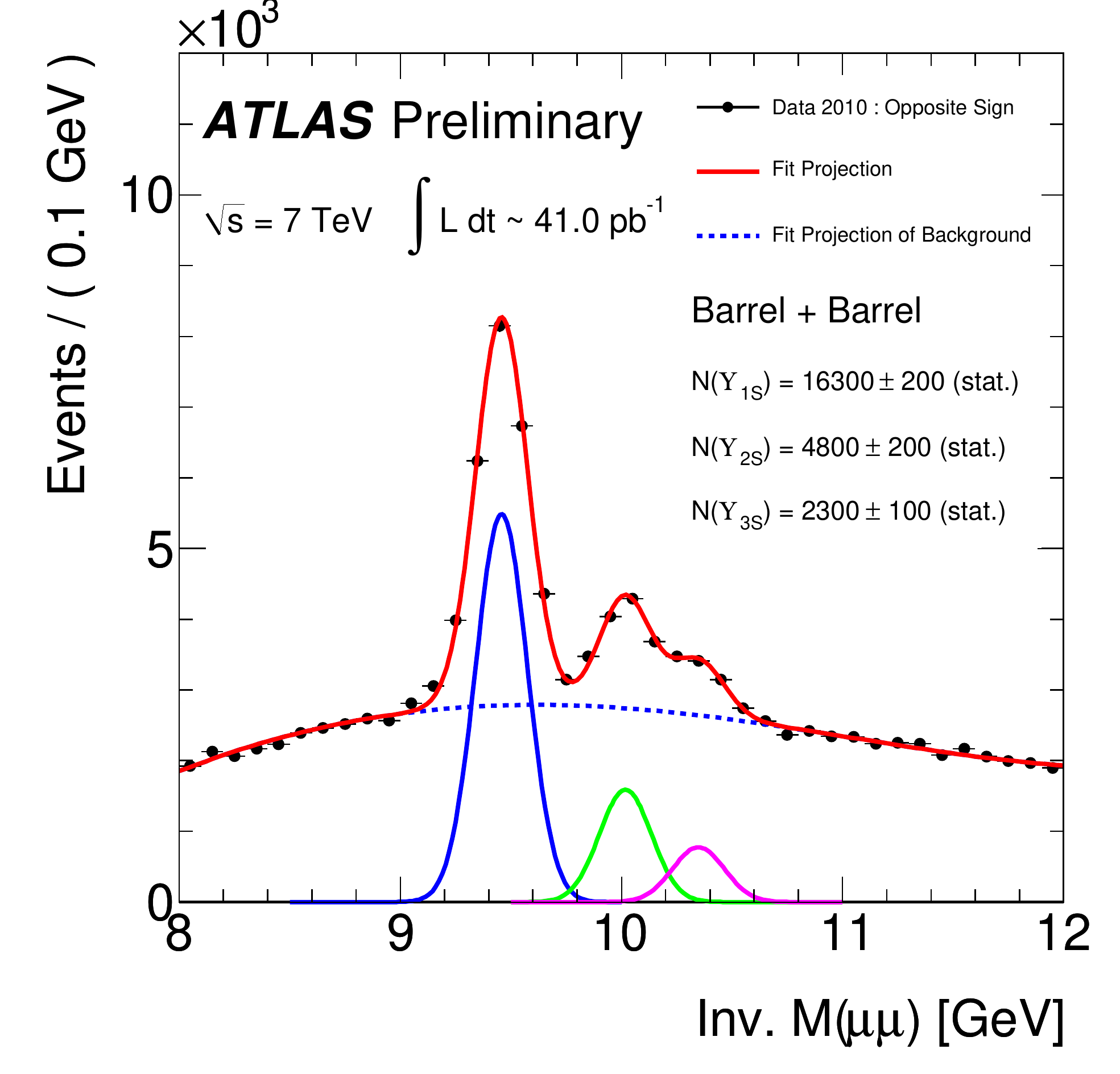}
\includegraphics[height=0.38\textwidth]{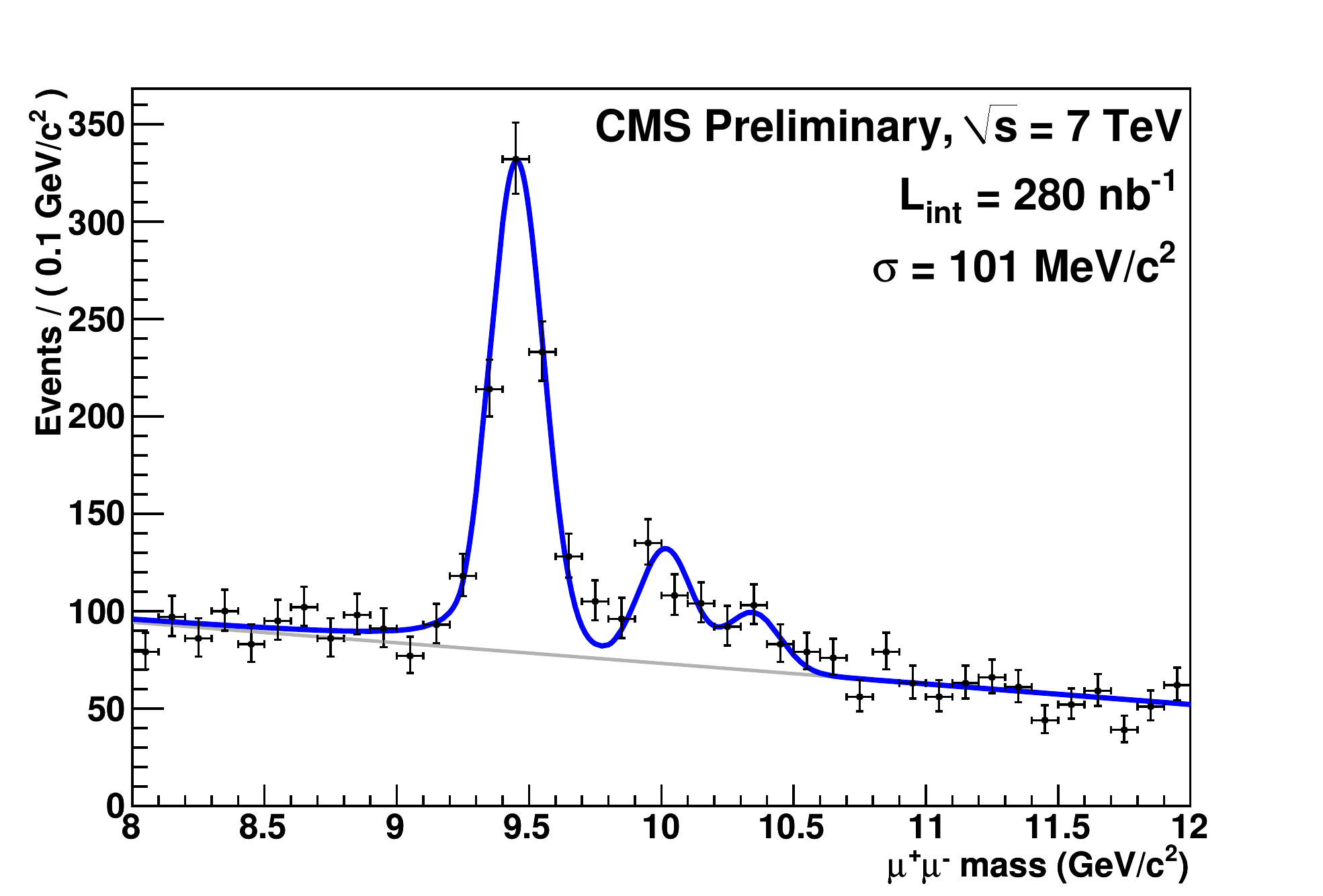}
\includegraphics[height=0.38\textwidth]{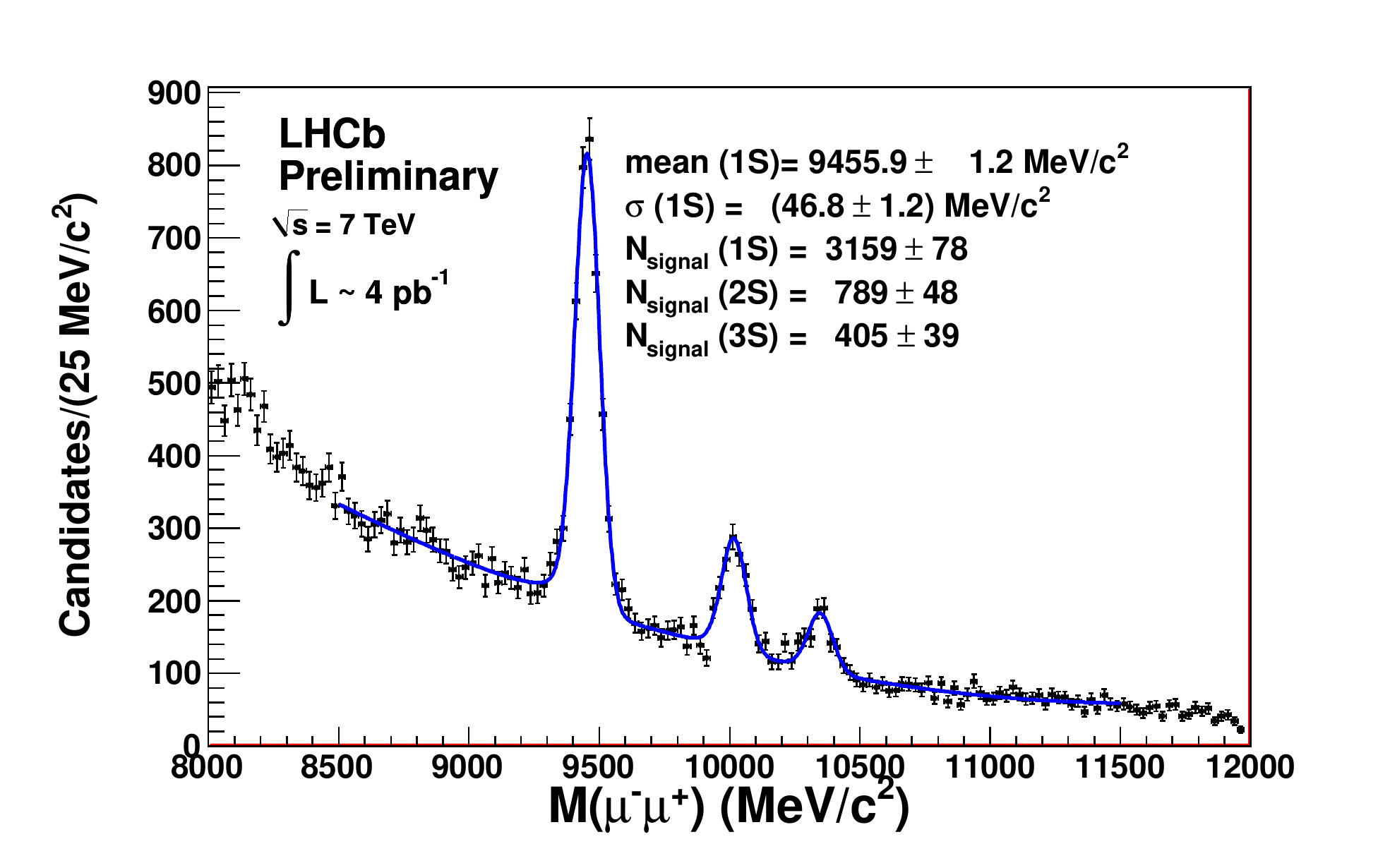}
\caption{The $\Upsilon$ family of resonances as reconstructed in the $\mu^+\mu^-$ final state.  Top left: ATLAS ($41 \, {\rm pb^{-1}}$); top right: CMS ($280 \, {\rm nb^{-1}}$)~\cite{CMSUPSILON_ICHEP}; bottom: LHCb ($\sim 4 \, {\rm pb^{-1}}$ ).}
\label{fig:upsilon}
\end{figure}

The onia programme of the LHC experiments has been extended to studies of the $b\bar{b}$ system.
In Fig.~\ref{fig:upsilon} are shown the $\Upsilon$ family of resonances from ATLAS, CMS and LHCb as
reconstructed in the $\mu^+\mu^-$ final state.  These plots provide a good benchmark to assess
the relative mass-resolution performances of the three experiments.  CMS has also used these data
to perform cross-section measurements~\cite{CMS_UPSILON}.

\begin{figure}[htb]
\centering
\includegraphics[height=0.30\textwidth]{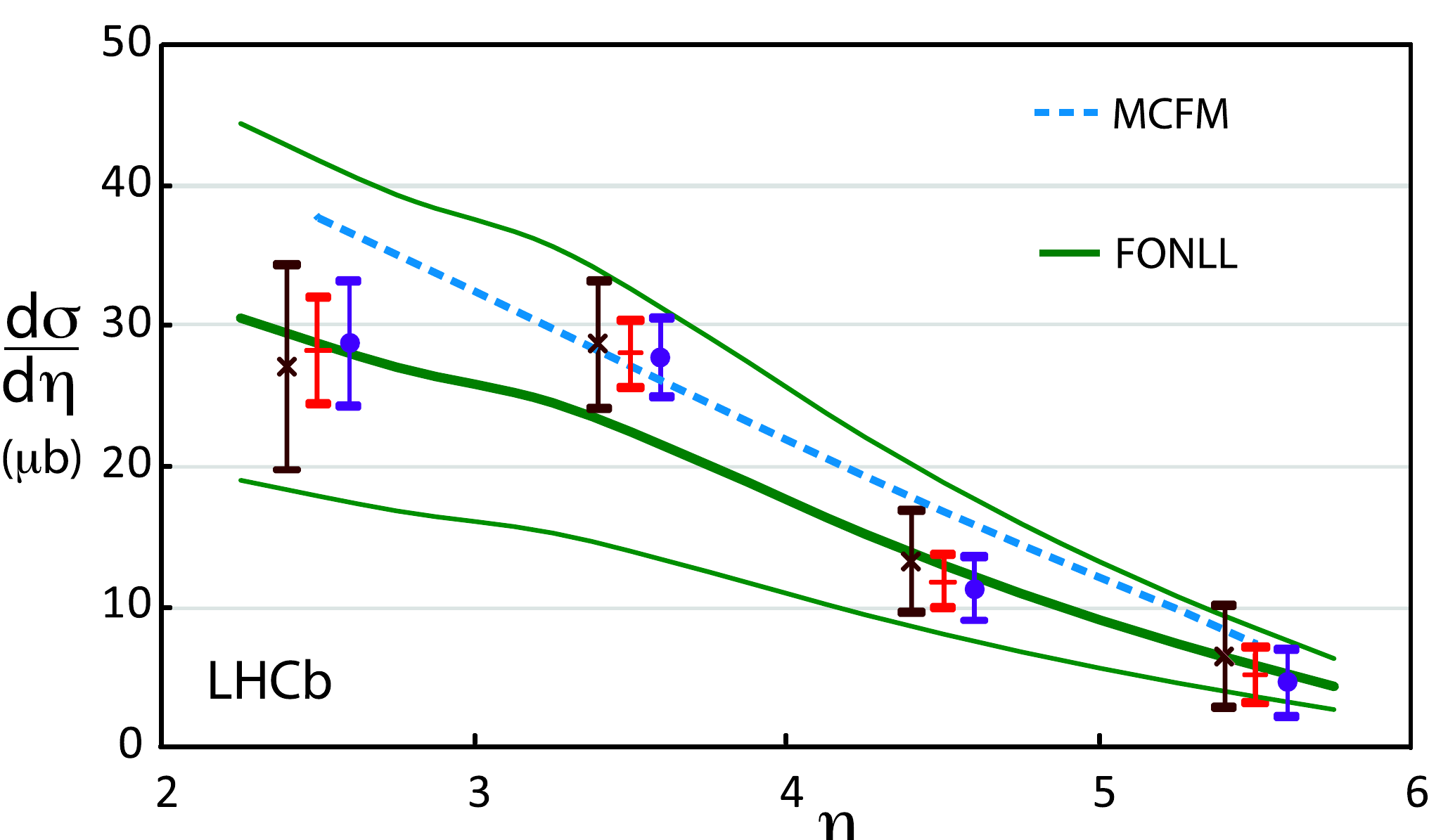}
\includegraphics[height=0.32\textwidth]{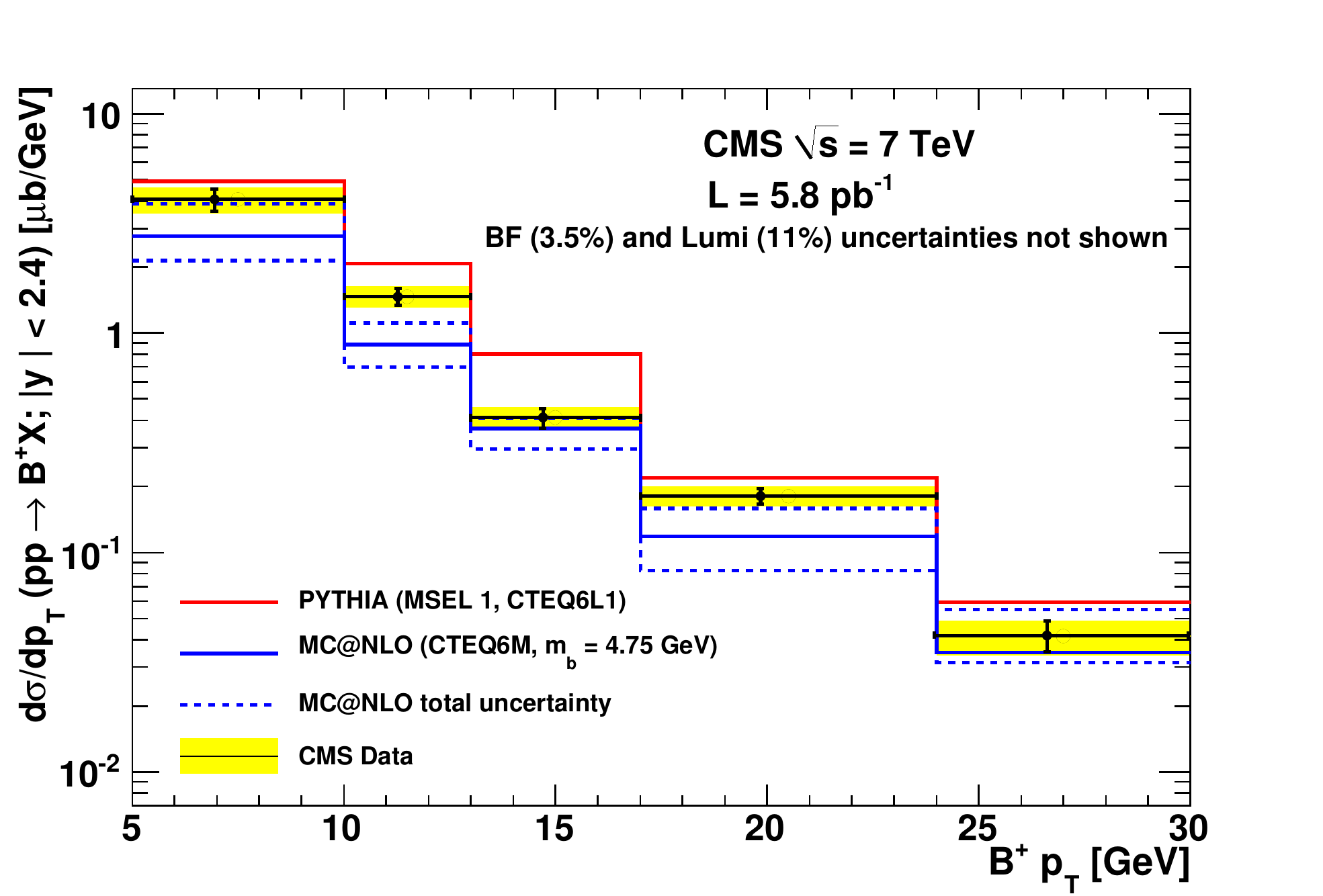}
\caption{Left: LHCb measured production cross-section at $\sqrt{s}=$~7 TeV as a function of pseudo-rapidity averaged over $b$-flavoured and $\bar{b}$-flavoured hadrons (the middle point shown in each bin is the average of the other two points, which come from separate trigger samples)~\cite{LHCB_XSEC}.
Right: CMS measured differential cross-section at $\sqrt{s}=$~7~TeV for $B^+$ production, with respect to transverse momentum~\cite{CMS_BPLUS}. 
Details on the theory predictions can be found in the references.}
\label{fig:bxsec}
\end{figure}

First measurements have been made of the production cross-section of $b$-flavoured hadrons at the LHC.
An early publication by LHCb employed the signature of  $D^0$-mesons displaced from
the primary vertex along with a muon with the correct sign correlation for both to arise from
semileptonic $b$-hadron decays~\cite{LHCB_XSEC}.  More recently, CMS has studied the production
of $B^+$ mesons through the decay $B^+ \to  J/\psi K^+$~\cite{CMS_BPLUS},
and also inclusive $b$-hadron production through the identification of muons with significant transverse momentum
with respect to the closest lying jet~\cite{CMS_XSEC}.  The LHCb measurements
are in good agreement with the theory predictions with which they are compared (see Fig.~\ref{fig:bxsec} left),
whereas the CMS measurements agree in shape with the predictions, but with a normalisation 
approximately 1.5 times larger than the MC@NLO expectation~\cite{MCNLO} (see Fig.~\ref{fig:bxsec} right).
These results confirm that the assumptions used for the $b\bar{b}$ cross-section in LHC flavour-physics sensitivity studies,
for example~\cite{ROADMAP},  were not overestimates.

The LHCb semi-leptonic analysis is being extended to look for  $D^0$, $D^+$, $D_s^+$ and $\Lambda_c$ decays in conjunction with
a lepton.  When cross-feed is accounted for, it then becomes possible to calculate the relative contribution of each $b$-hadron species.
A preliminary result has been determined for $f_s/(f_u+f_d)$, the fraction of $B^0_s$ to 
the sum of $B^0$ and $B^+$ hadrons at the LHC at $\sqrt{s}=7$~TeV in the forward region~\cite{LHCBSEMILEP_DEC}: 
$ f_s/(f_u+f_d) = 0.130 \pm 0.004 \,({\rm stat}) \pm 0.013\, ({\rm syst.}).$

Another interesting topic of study in heavy flavour production is that of the correlation between the kinematical properties of 
the $b$- and $\bar{b}$-hadrons.  
CMS has made the first analysis of the angular correlations between $b$- and $\bar{b}$-hadrons
at the LHC using a secondary vertex reconstruction method~\cite{CMSCORRELATION}.  It is found that a
sizable fraction of the beauty hadron pairs are produced with small opening angles.
 
LHCb has also performed a preliminary measurement of the cross-section of $D^0$, $D^+$ and $D^+_s$ production
within its acceptance~\cite{LHCB_CHARM}.  It is found that at $\sqrt{s}=7$~TeV
$c\bar{c}$ production is around twenty times more abundant than that of $b\bar{b}$ events.
Again these results are found to be in good agreement with QCD
expectation.

\section{Expectations over the coming one-to-two years}

In the following few pages some examples are given of topics where interesting 
new results and updates are soon to be expected.  This list is selective 
and many important analyses (for example the forward-backward
asymmetry in $B^0 \to K^{*}\mu^+\mu^-$ decays) are not discussed.

\subsection{{\boldmath $B_s^0 \to \mu^+ \mu^-$}}

The channel $B_s^0 \to \mu^+\mu^-$ is the $b$-physics rare decay {\it par excellence}.
In the Standard Model (SM) the predicted branching fraction for this mode, at $(3.35 \pm 0.32) \times 10^{-9}$~\cite{BLANKE}, 
is both exceedingly low, and precisely determined.  In many New Physics (NP) models, however, significant enhancements are possible.
In particular, in the CMSSM at large $\tan \beta$, the branching fraction goes as $\tan^6 \beta$, meaning
that searches for this mode have great discovery potential; conversely, knowing that the branching 
ratio of the decay lies below a certain value can  impose severe constraints on  NP parameter space~\cite{ELLIS}.

The present 90\% C.L. upper limits on this decay are $36 \times 10^{-9}$ from CDF with $3.7 \,{\rm fb^{-1}}$  of 
data~\cite{CDF_BSTOMUMU}, and  $42 \times 10^{-9}$ from D0 with $6.1 \,{\rm fb^{-1}}$ of data~\cite{D0_BSTOMUMU}.
Updates with the full Run II dataset will allow for improved sensitivity~\cite{BERTRAM},  but it is at the LHC where there are 
the highest hopes of observing this decay~\cite{SERRA}.   

The $B_s^0 \to \mu^+\mu^-$ search is well suited to both ATLAS and CMS~\cite{CMS_BSTOMUMU},
on account of a decay signature that allows for a high trigger efficiency, and can be pursued even when the accelerator 
is operating at a luminosity of $10^{34} \,{\rm cm^{-2} s^{-1}}$.  LHCb also has very good prospects.
When performing this search, the sample is typically dominated after pre-selection by
events containing two genuine muons, though not arising from the same decay vertex.
For this reason topological information concerning the quality and isolation of the decay
vertex, and the kinematics of the candidate is critical.
LHCb will be able to make use of a  clean sample of $B^0, B^0_s \to h^+h'^-$  ($h,h'=\pi,K$) events 
on which to calibrate the topological and kinematical variables that are used in the discrimination of
signal from background.  Shown in Fig.~\ref{fig:lhcbbstomumu} is the expected performance of LHCb
as a function of integrated luminosity, in terms of exclusion and observation.  These curves are 
determined from Monte Carlo studies, but assume the measured $b\bar{b}$ cross-section at $\sqrt{s}=7$~TeV.
Studies with early data indicate that the Monte Carlo predictions are realistic.  With the $\sim 1$~{$\rm fb^{-1}$  of
data expected in 2011 the $90\%$ exclusion limit should drop below $10^{-8}$, or indeed evidence 
of a non-SM signal may emerge.  A five sigma observation of  the signal at the branching ratio
predicted in the SM will be possible, but will require several years of operation.

\begin{figure}[htb]
\centering
\includegraphics[height=0.39\textwidth]{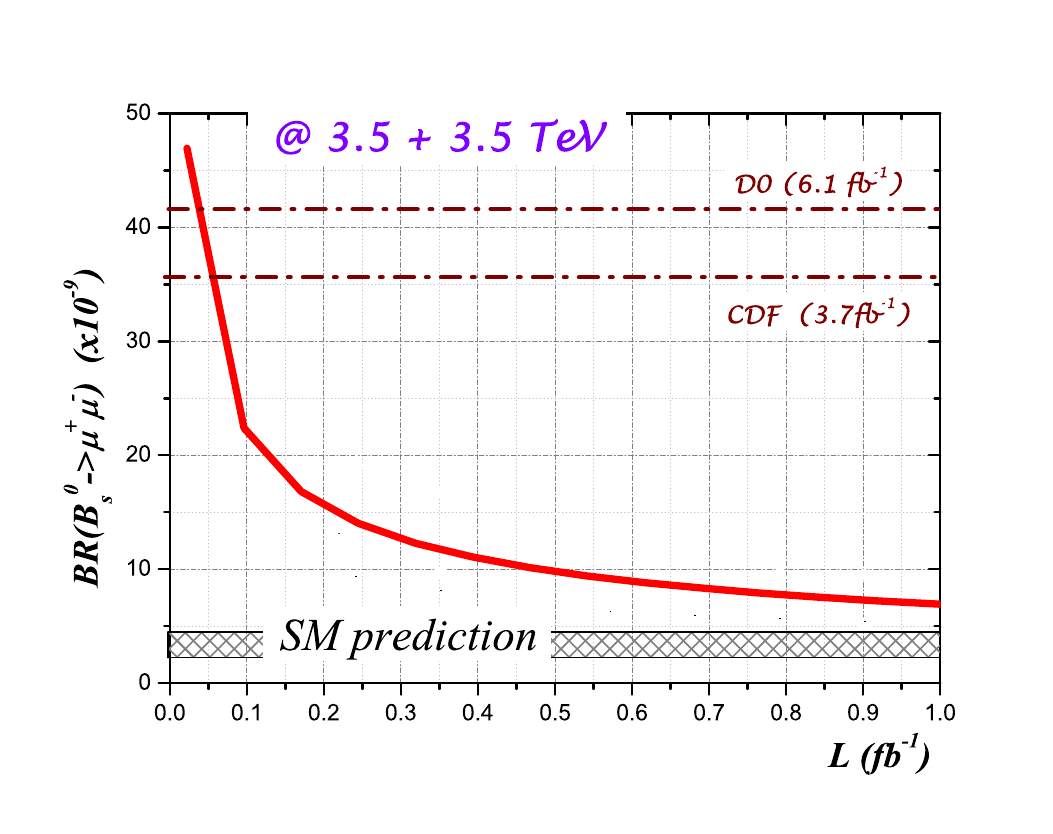} 
\includegraphics[height=0.39\textwidth]{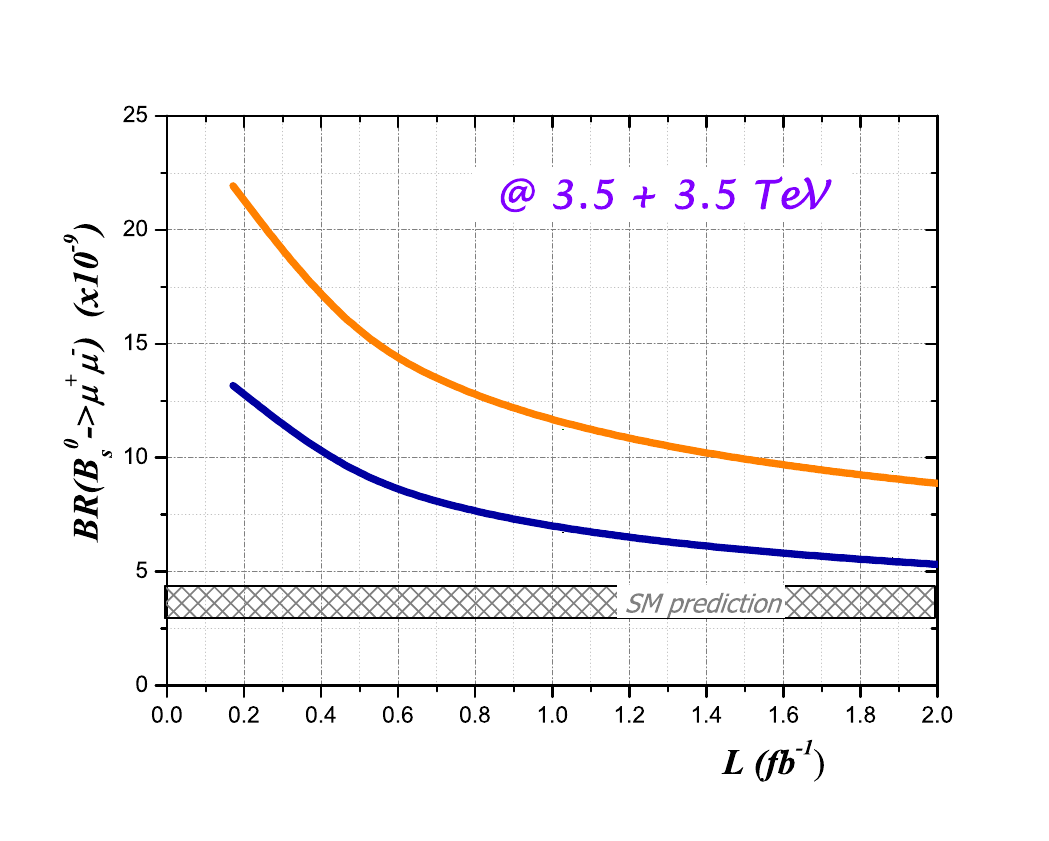} 
\caption{LHCb expected performance in the search for $B^0_s \to \mu^+\mu^-$ as a function
of integrated luminosity at $\sqrt{s}=7$~TeV.  Left: 90\% exclusion.
Right: discovery, with the upper and lower curves representing 5 and 3 $\sigma$ respectively.}
\label{fig:lhcbbstomumu}
\end{figure}

\subsection{CP violation in the {\boldmath $B^0_s-\bar{B^0_s}$} system}

Mixing-induced CP violation has not yet been observed in the
$B^0_s-\bar{B^0_s}$ system.  This is because it is predicted to be small in the SM,
and also because only recently has the Tevatron acquired the necessary
statistics to embark on a meaningful measurement programme.
(Although Belle took $B^0_s$ data at the $\Upsilon(5S)$,
the boost was inadequate to resolve
the very rapid oscillations.)

The preferred channel for studying CP-violation in the interference
between $B^0_s$ mixing and decay is $B^0 \to J/\psi \phi$. 
In the SM the predicted value of the phase, $2 \beta_s$, probed by this decay
is very small and tightly constrained: $2 \beta_s = -0.0366 \pm 0.0014$~\cite{CKMFITTER}.
The most recent preliminary results available from CDF~\cite{CDFPHIS,KREPS} and D0~\cite{D0PHIS,GUENNADI},
performed with 6.1~$\rm {fb^{-1}}$ and 5.2~$\rm fb^{-1}$ of data respectively, have a precision
which is an order of magnitude too poor to be sensitive to CP-violation at the SM level.
Nevertheless, NP enhancements could generate a much larger value,
and indeed both measurements give weak hints, at the one sigma level, of such
a possibility. 

Whether there is NP affecting $\beta_s$ at such an enhanced level
should soon be clear at the LHC.  LHCb, in particular, is well suited to this
measurement~\cite{STEPHIE}; simulation studies indicate that if the true phase is at the
value suggested by the Tevatron results then five sigma observation will
be possible with 0.1-0.2~$\rm fb^{-1}$ of data.
The first steps on the road to performing this measurement have already
been taken in 2010, with a clean sample of $B^0_s \to J/\psi \phi$ 
events accumulated (see Fig.~\ref{fig:jpsix}, left),
and a proper time resolution that is found to be $\sim 50$~$\rm {fs}$, which
is already close to Monte Carlo expectations.  Furthermore, other 
decay modes are now under consideration to augment the $\beta_s$ sensitivity.
In particular, the decay $B^0_s \to J/\psi f_0(980)$ is a very attractive
possibility, being a CP-eigenstate which therefore does not require
an angular analysis, in contrast to  the case of  $B^0_s \to J/\psi \phi$.
This decay has been observed for the first time in the 2010 LHC
run (see Fig.~\ref{fig:jpsix}, right)~\cite{LHCBJPSIF0} with a rate
that indeed makes it useful for the $\beta_s$ measurement.

Another interesting quantity that exists in the $B^0_s - \bar{B^0_s}$ system
is the flavour-specific asymmetry, which may be measured in semi-leptonic
decays, and which is sensitive to CP-violation in  $B^0_s - \bar{B^0_s}$
mixing.  
D0 has recently released a result~\cite{GUENNADI,D0ASL}, based on 6.1~$\rm {fb^{-1}}$,
for an observable which is
effectively the sum of the flavour asymmetries in $B^0$ and $B^0_s$ mesons.
Very intriguingly the measurement yields an asymmetry of order 1\%,
which though small is still two orders of magnitude, and 3 measurement
sigma, bigger than the tiny effect expected in the SM.
Updates  are awaited with great interest, both from the Tevatron
and the LHC.  In this measurement systematic control is crucial --
percent level biases can easily enter from background asymmetries and
detector effects.  The LHC has the additional challenge of combating 
production asymmetries which are in general non-zero on account of
the $pp$ initial state.

\begin{figure}[htb]
\centering
\includegraphics[height=0.34\textwidth]{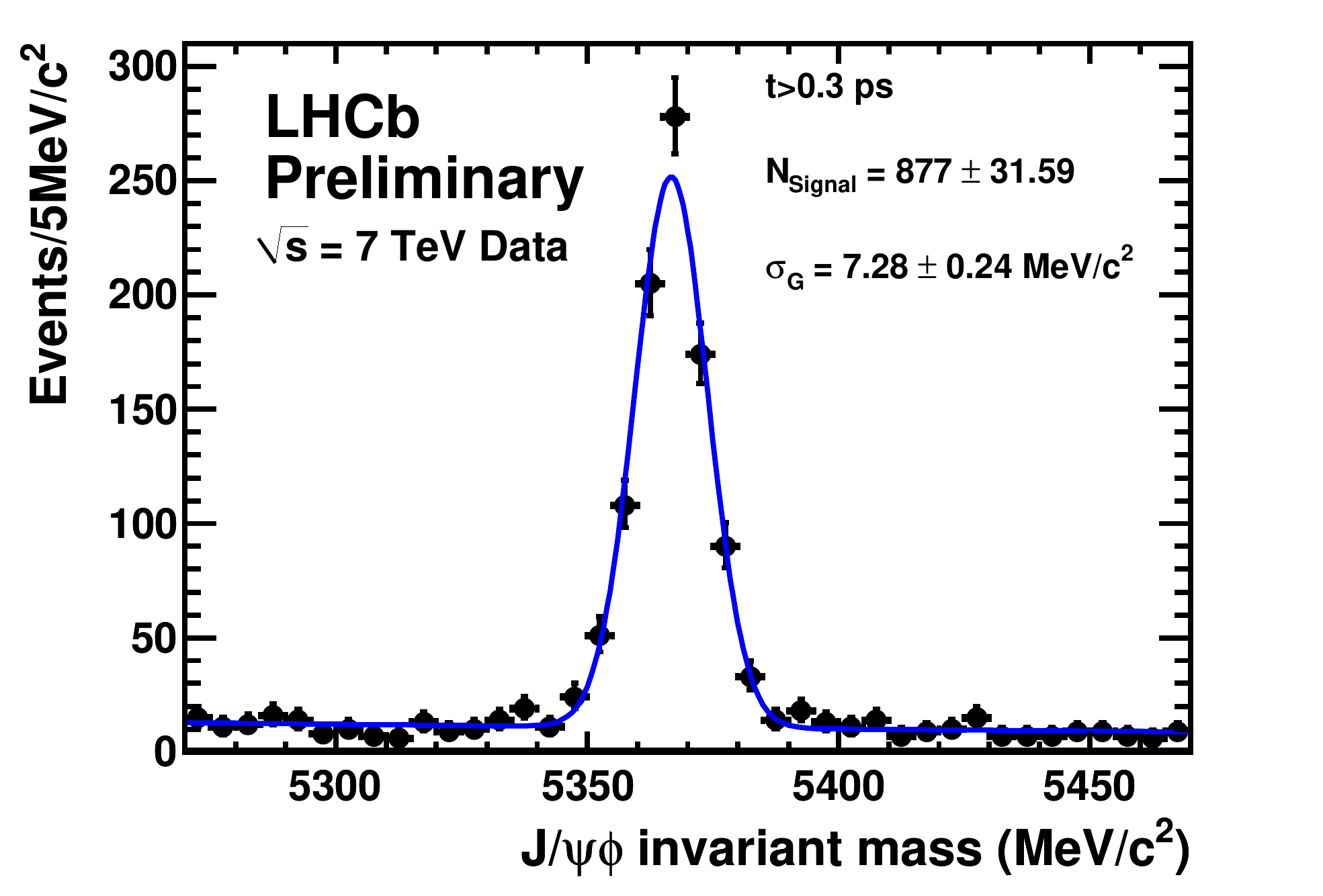} 
\includegraphics[height=0.33\textwidth]{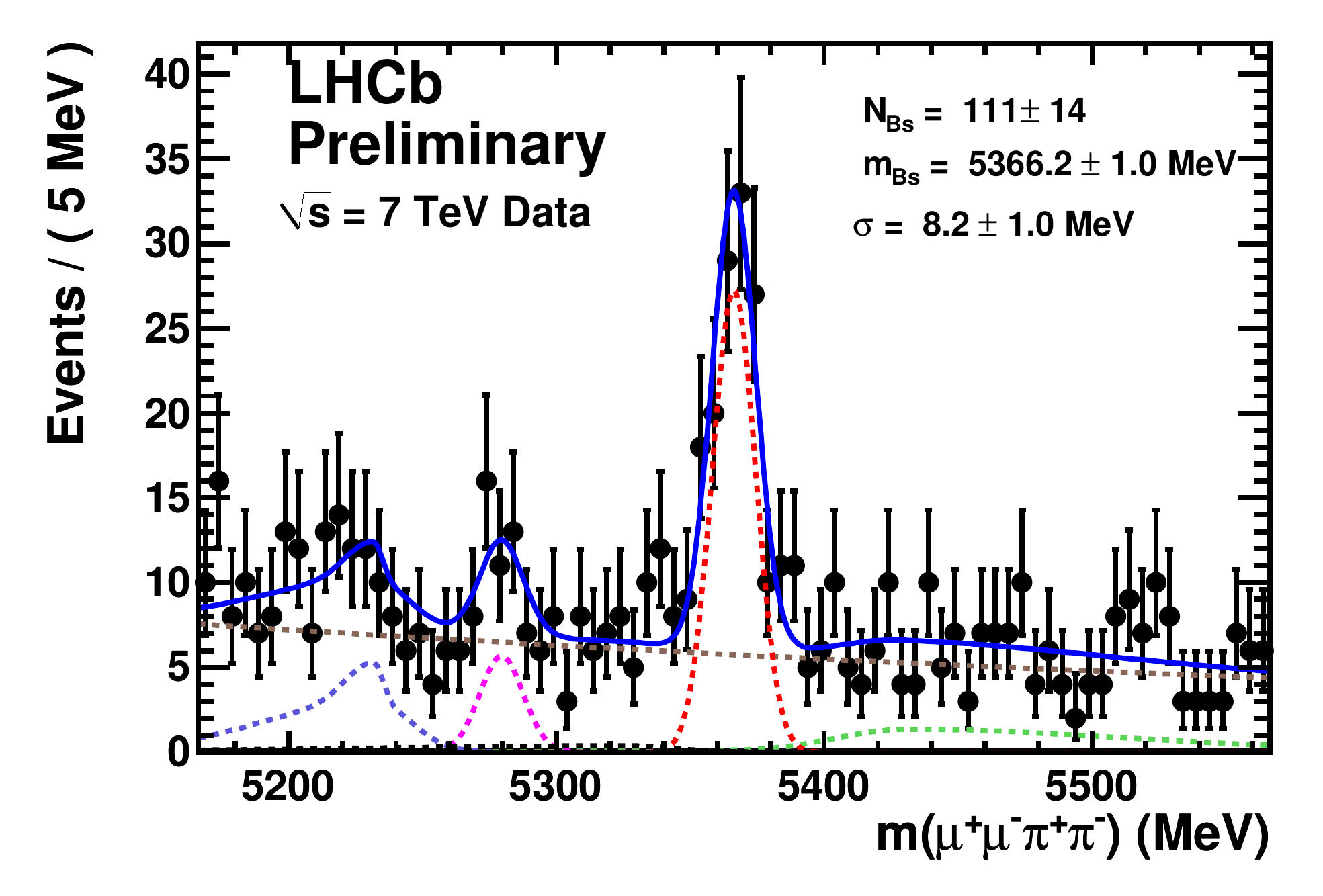} 
\caption{Left: LHCb $B^0_s \to J/\psi \phi$ with 34~$\rm pb^{-1}$. Right: LHCb first observation of $B^0_s \to J/\psi f_0(980)$ performed with 33~$\rm pb^{-1}$~\cite{LHCBJPSIF0}.}
\label{fig:jpsix}
\end{figure}

\subsection{Hadronic ${\boldmath b}$-decays}

The studies considered so far all involve channels with two leptons in the final state.
This  characteristic provides a very distinctive trigger signature which can be exploited
by any general purpose detector at a hadron collider.  Attaining good efficiency
on heavy mesons decays into hadronic final states, on the other hand, requires 
a more specialised trigger strategy.  The track trigger of CDF and the high-$p_T$ hadron
trigger of LHCb are two examples of trigger systems which can perform this task.

Perhaps the most important physics goal in hadronic $b$-decays is the
determination of the unitarity triangle angle $\gamma$.   Provided that
the trigger of the experiment  has sufficient efficiency for the decays of interest, 
this measurement  is very suited to hadron colliders. 
Firstly, the sample sizes that could be collected at the $B$-factories were 
inadequate to allow for a measurement with a precision better than 10$^\circ$.  The statistics that
will be collected at LHCb, in particular, will be much larger.
Secondly, the very powerful
suite of measurements of the sort $B^\pm \to D K^\pm$ do not
require a time-dependent analysis and have no need of
flavour-tagging.  (Flavour tagging is a priori more powerful at the $\Upsilon(4S)$
than at a hadron collider because of the quantum-correlations between
the produced $B$-mesons.)  
Finally, experiments at hadron colliders
have the attractive possibility of exploiting strategies involving $B^0_s$
mesons, such as the time-dependent study of the decay $B^0_s \to D_s K$. 

CDF has already demonstrated its capabilities in $B \to DK$ studies, by performing
a first `GLW'~\cite{GLW} $B^\pm \to D(KK,\pi\pi) h^\pm$  ($h=\pi,K$) analysis~\cite{CDF_GLW}
with around 1~$\rm fb^{-1}$ of data.  At this conference an `ADS'~\cite{ADS} $B^\pm \to D(K\pi) h^\pm$  
($h=\pi,K$) study, based on 5~$\rm fb^{-1}$, was presented for the first  time~\cite{CDF_ADS} (see Fig.~\ref{fig:cdf_ads}).
These analyses cannot be used in isolation to extract
$\gamma$, but they provide useful input to the global picture and make
clear the potential of hadronic experiments in this field.

\begin{figure}[htb]
\centering
\includegraphics[height=0.40\textwidth]{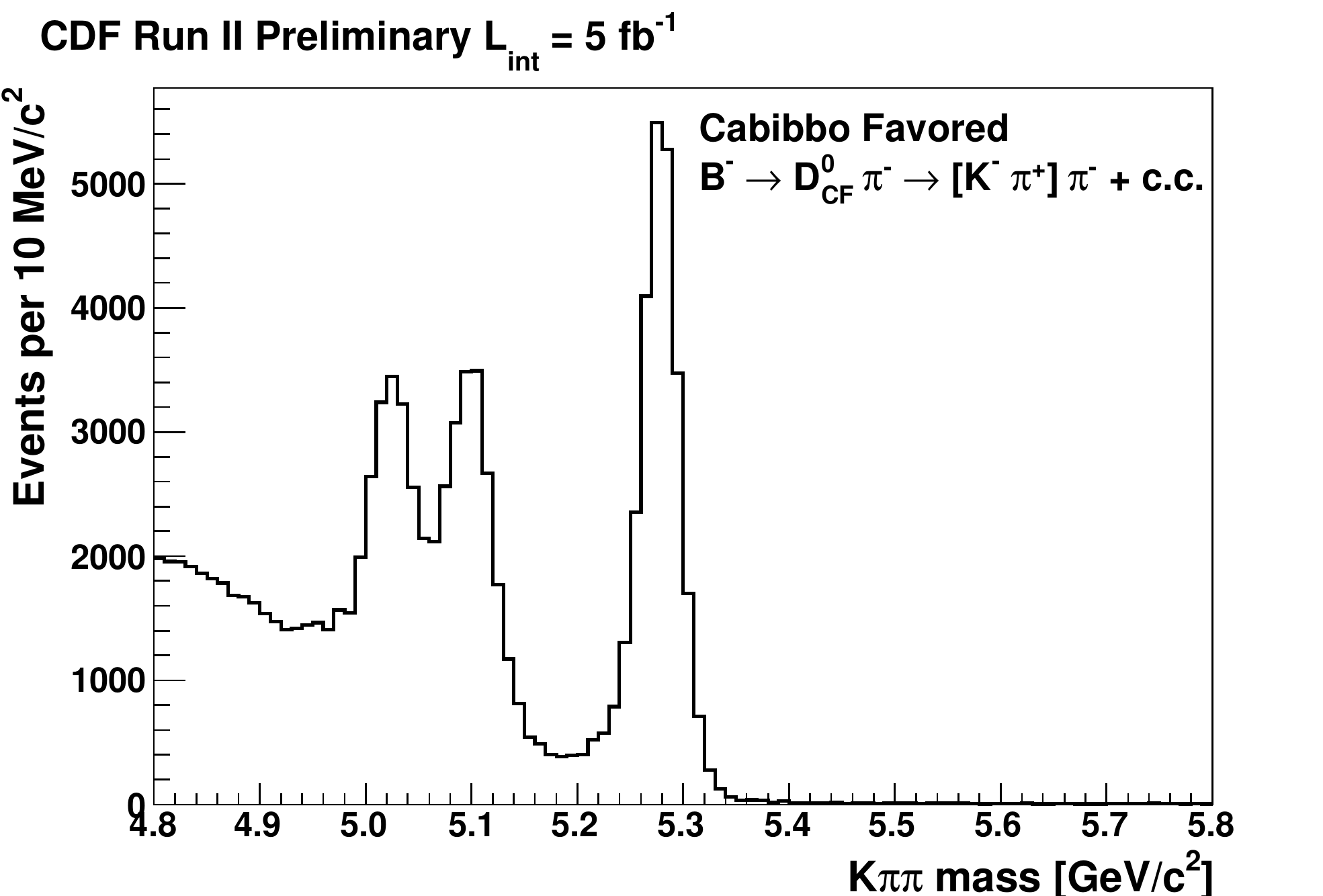} 
\caption{CDF Cabibbo favoured $B^- \to D^0(K\pi) \pi^-$ events collected with 5~fb$^{-1}$~\cite{CDF_ADS}.}
\label{fig:cdf_ads}
\end{figure}

LHCb has already accumulated significant and clean samples in $B \to Dh$ ($h= \pi, K$) decays.  
As well as benefiting from its
efficient trigger for hadronic decays, LHCb also profits from using its Ring Imaging Cherenkov Counter (RICH) system to
distinguish between pions and kaons.  This is illustrated in Fig.~\ref{fig:lhcbbtodh}, where
the RICH is used to distinguish between $B \to D \pi$ and $B \to D K$ decays.
Already with a 1~$\rm fb^{-1}$ dataset it 
should prove possible to improve on many of the measurements performed at the $B$-factories~\cite{VAVAMIKE}.

\begin{figure}[htb]
\centering
\includegraphics[height=0.34\textwidth]{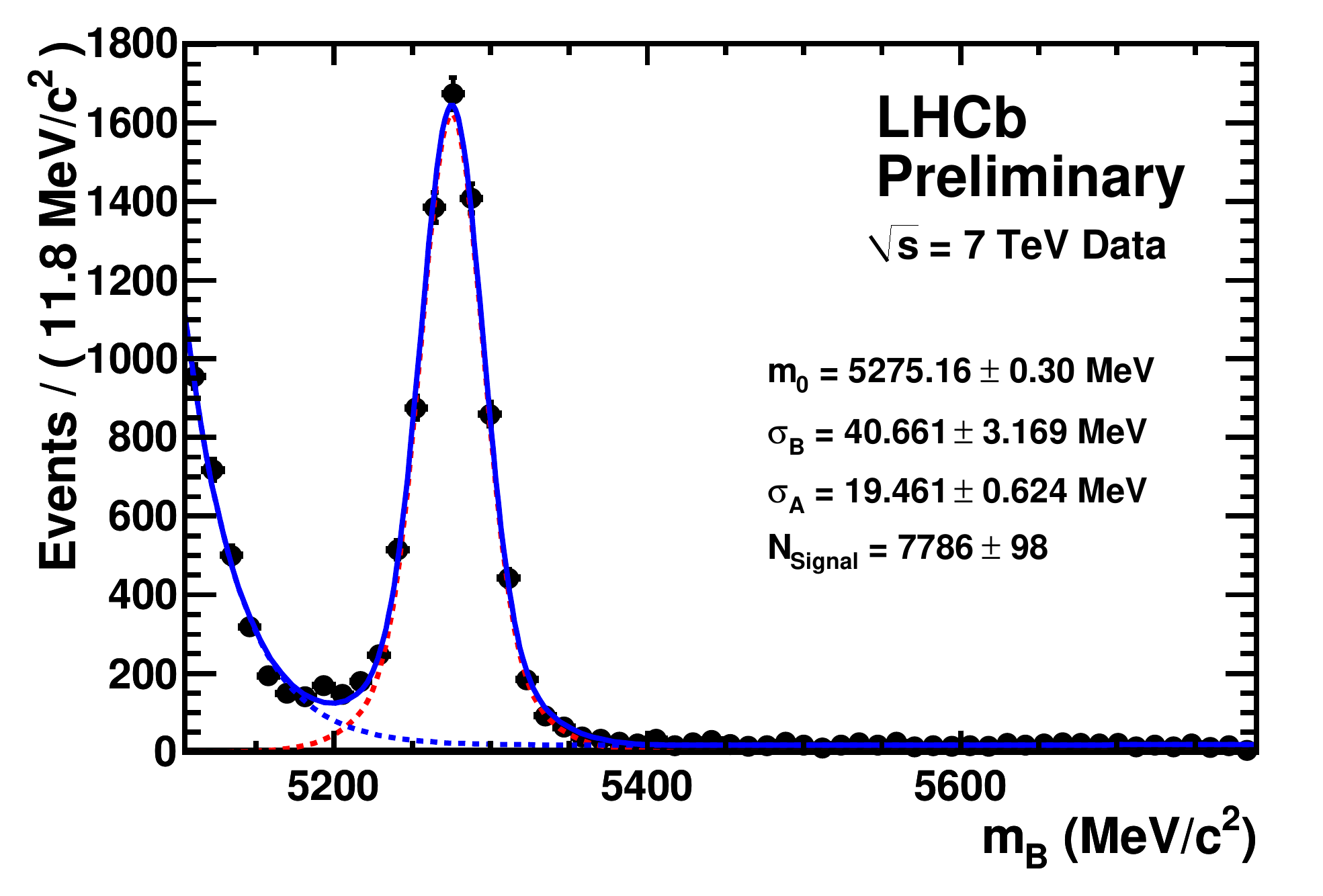} 
\includegraphics[height=0.33\textwidth]{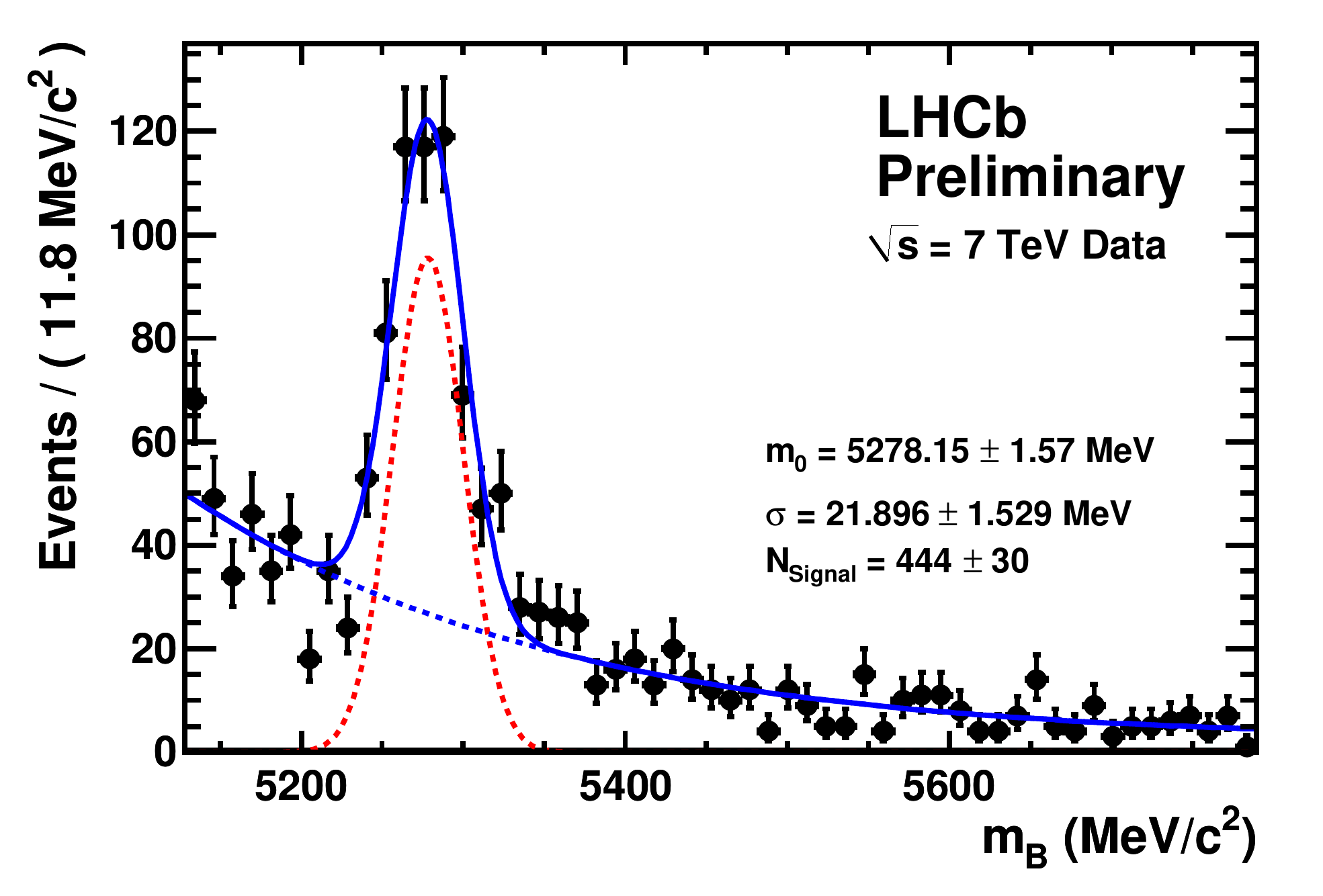}  
\caption{LHCb $B \to D (K\pi) h$ decays with 34~$\rm pb^{-1}$. Left: using RICH to select $B \to D \pi$.  Right: using RICH to select $B \to DK$.
(Note that the plots cannot be interpreted in a quantitative manner without  knowledge of the RICH particle identification performance.)}
\label{fig:lhcbbtodh}
\end{figure}

\subsection{Charm physics}

The observation of charm mixing has been one of the most interesting
discoveries in particle physics in recent years.   Although the $B$-factories
took centre stage in this discovery, it must be remembered that CDF
played its part, with a high sensitivity study of `wrong sign' $D^0 \to K\pi$ 
decays~\cite{CDF_DMIX}.  The priority is now to improve the sensitivity
of such measurements, and those of the time-integrated and time-independent studies, to search for 
CP-violation in the charm system.  It is clear that over the 
coming few years the precision of these measurements will improve
greatly, given the enormous statistics still to be exploited at
the Tevatron, and foreseen at LHCb~\cite{GERSABECK}.  Indeed,  one  month after 
the end of the workshop, the preliminary result of a 6~$\rm fb^{-1}$  study of the time
integrated CP-asymmetry in $D^0 \to \pi\pi$ events was made public~\cite{CDF_DCP}.
When interpreted as a search for direct CP-violation, this measurement 
has around twice the precision of those performed at BABAR and Belle. 

\section{Conclusions}

For several years at least (and with the exception of BES-III), $B$ and $D$ physics
will be pursued solely at hadron colliders.  Although the $\Upsilon(4S)$ is a wonderful
environment for flavour studies,  the power and potential of the data still to
be exploited at CDF and D0,  and those now accumulating at LHCb, is difficult to overstate.
The $B^0_s$ meson and $b$-baryon sectors will start to reveal their secrets, and very high statistics studies
will continue with $B^0$, $B^+$ and $D$ mesons.  It is certainly no dark age for flavour studies -- here comes the sun!


\Acknowledgements
I am grateful to the organisers for arranging a most stimulating meeting.
I also thank Tim Gershon, Giovanni Punzi, Olivier Schneider for their careful reading of the first
draft of these proceedings.  I am responsible for any errors that remain.

\end{document}